\documentclass[amsmath,amssymb]{revtex4}

\usepackage{graphicx}
\usepackage{dcolumn}
\usepackage{bm}
\usepackage{here}
\newcommand{\vect}[1]{\mbox{\boldmath $#1$}}
\begin{document}

\title[Dependence of self-assembled amphiphile structure]
{Dependence of self-assembled amphiphile structure on interaction between
hydrophilic groups}

\author{HIROAKI NAKAMURA} 
 \email{nakamura@tcsc.nifs.ac.jp}
\author{YUICHI TAMURA}
\affiliation{National Institute for Fusion Science, 
        Oroshi-cho 322-6, Toki 509-5292, JAPAN}
 
\date{23 August 2005 and accepted 12 April 2006}


\begin{abstract}
In a previous study ({\it Comp. Phys. Com.\/} (2005) {\bf 169}, 139--143), we clarified the dependence of the phase structure on the hydrophilicity of an amphiphilic molecule
by varying the interaction potential between the hydrophilic molecule and water ($a_{\rm AW}$) in a dissipative particle dynamics (DPD) simulation using the Jury model.
In the present paper, we perform another DPD simulation using the previous model to investigate the dependence of the interaction potential between adjacent hydrophilic groups on the phase structure.
By varying the coefficient of the interaction potential between adjacent hydrophilic groups $a_{\rm AA}$ ($a_{\rm AA}=15,25,40,$ and $250$) at a dimensionless temperature of $T=0.5$ and a concentration of amphiphilic molecules in water of $\phi=50\%,$  hexagonal ($a_{\rm AA}=14,25,40$) and micellar ($a_{\rm AA}=250$) phases were observed. 
In comparison with the previous results, the dependence of the A-B dimer's shape on $a_{\rm AA}$ was determined to be weaker than that on $a_{\rm AW}$.
Therefore, it is concluded that the solvent water $W$ plays an important role in aggregation of the A-B dimers.
\end{abstract}

\maketitle
\section{Introduction}
Amphiphilic molecules have many degrees of freedom in the structures they adopt.  
However, when the temperature and concentration of an aqueous solution of the amphiphilic molecules are fixed, the molecules are restricted to a certain molecular shape and aggregate to form a variety of mesoscopic structures, for example, micellar, lamellar and hexagonal phases.

In previous papers\cite{04Nakamura,05Nakamura}, we have investigated the amphiphilic molecule hexaethylene glycol dodecyl ether (C$_{12}$E$_{6}$),  a popular surfactant in water that forms a variety of self-assembled structures.
The phase structure of C$_{12}$E$_{6}$ was investigated by Mitchell\cite{83Mitchell} in 1983.
In our work, we clarified\cite{05Nakamura} the dependence of the phase structure on hydrophilicity by varying the interaction potential ($a_{\rm AW}$) between the hydrophilic molecules and water in a dissipative particle dynamics (DPD) simulation using the Jury model\cite{99Jury}.

In the present paper, to determine the relative dominance of different microscopic interaction potentials in determining the mesoscopic phase structure, we investigated the dependence of the phase structure on the interaction potential between adjacent hydrophilic groups $a_{\rm AA}.$
The details of the DPD simulation algorithm and models can not be included due to lack of space but are the same as reported previously\cite{05Nakamura} with the exception of the interaction potentials $a_{\rm AW}$ and $a_{\rm AA}$. 
The differences with the previous simulation are described in Section 2.

\section{Simulation Method}
We used the same DPD model and algorithm\cite{99Jury,97Groot} as described previously\cite{04Nakamura, 05Nakamura}, with a modification of the conservative force $\vect{F}^{\rm C}_{ij}$ between particles $i$ and $j$, given in the present paper by
\begin{equation}
\vect{F}^{\rm C}_{ij} \equiv   
  \left\{
    \begin{array}{ll}
      a_{i j} (1-r_{i j}) \vect{n}_{ij} 
         & \mbox{if  }  r_{i j}<1,\\
    0 & \mbox{if  } r_{i j} \ge 1,     
    \end{array}
  \right.
\label{eq.fc}
\end{equation}
where $\vect{n}_{ij} \equiv ( \vect{r}_{i}-\vect{r}_{j} ) / |\vect{r}_{i}-\vect{r}_{j}|$, and $\vect{r}_{i}$ is a position vector for particle $i$.
Coefficients $a_{ij}$ in Eq. (\ref{eq.fc}) denote the coupling constants between particles $i$ and $j$.
The numerical values of $a_{ij}$ are given by Table \ref{aij}.


\section{Simulation Results and Discussion}\label{results}
To demonstrate the dependence of the mesoscopic phase structure on the interaction potential between adjacent hydrophilic groups, we varied the A-A interaction potential coefficient such that $a_{\rm AA}=15,25,40,$ and $250.$

\begin{figure}
\begin{center}
\includegraphics[width=12cm]{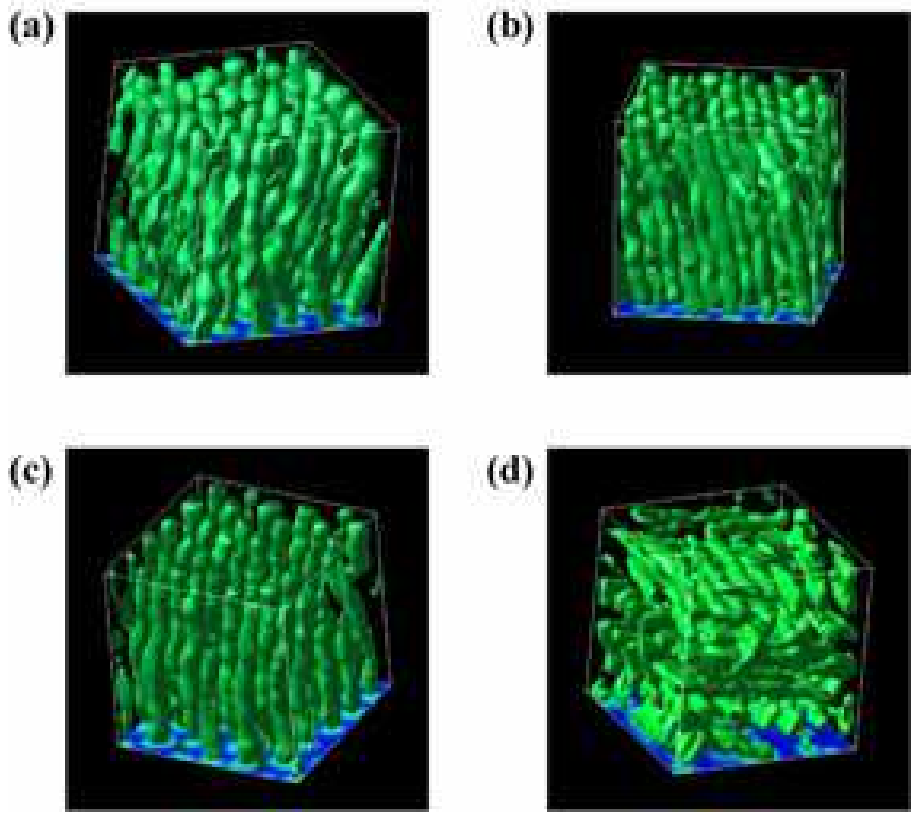}
\caption{\label{st1}Structures formed for each potential coefficient,
$a_{\rm AA}=15 (a), 25 (b), 40 (c),$ and $250 (d).$ We set $T=0.5$ and $\phi=50\%$ during simulation.}
\end{center}
\end{figure}

\begin{figure}
\begin{center}
\includegraphics[width=10cm]{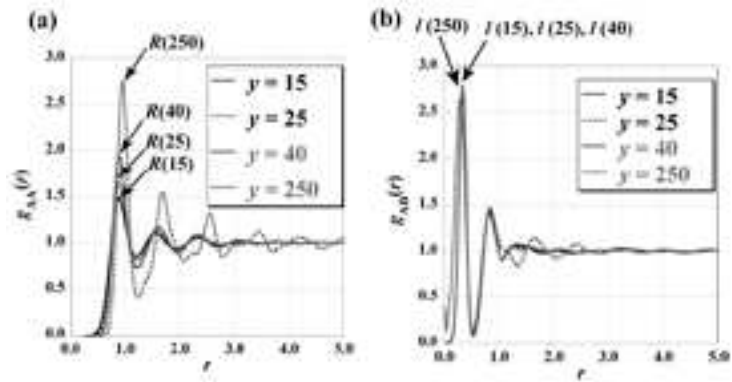}
\end{center}
\caption{\label{gr}Radial distribution functions for the solute particles vs. distance between the two particles $r$ for $a_{\rm AA}=15,25,40$ and $250$.
(a) The A-A radial distribution function $g_{\rm AA} (r)$. The first peak $R(a_{\rm AA})$ of each curve corresponds to the A-A distance between two adjacent AB dimers. 
$R(15)=0.889, R(25)=0.889, R(40)= 0.904, R(250)=0.963.$
(b) The A-B radial distribution function $g_{\rm AB}(r)$.
The first peaks correspond to the length of the A-B dimer $l$ in Fig.\ref{packing}.
 $l(15)=0.326, l(25)=0.326, l(40)= 0.326, l(250)=0.296.$
}
\end{figure}

The aggregation structure for each value of $a_{\rm AA}$ is shown in Fig. \ref{st1}.
To quantitatively classify the phase structure, we also plot the radial distribution function of the solute particles for each $a_{\rm AA}$ in Fig. \ref{gr}. 
The figures show that hexagonal phases ($a_{\rm AA}=15,25,40$) and micellar ($a_{\rm AA}=250$) phases are formed depending on the coefficient $a_{\rm AA}$ of the A-A interaction potential.

\begin{figure}[htbp]
\begin{center}
\includegraphics[width=10cm]{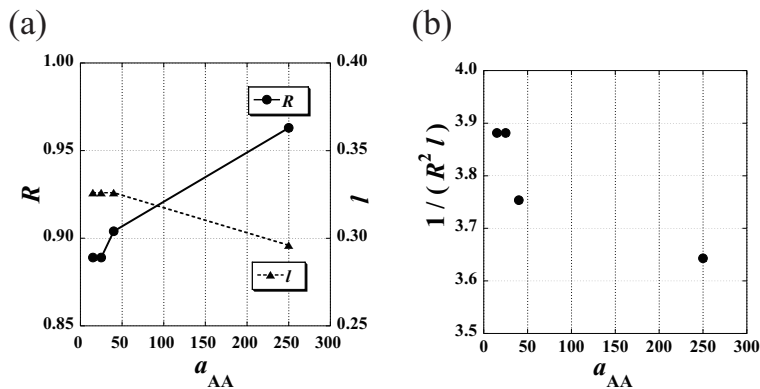}
\end{center}
\caption{ Shape parameters of two adjacent molecules vs. the A-A interaction potential coefficient $a_{\rm AA}$.
(a) The A-A distance between two adjacent AB dimers $R$ and the length of the hydrophobic tail  $l$ are shown.
(b) The inverse  of  $R^2 l $, which is proportional  to the packing parameter $p$, is plotted.
\label{ratio} }
\end{figure}

The distance $R$ between a hydrophilic group in an A-B dimer and another group in the nearest A-B dimer (Fig. \ref{packing}) is shown as the first peak $R(a_{\rm AA})$ of $g_{\rm AA}(r)$ in Fig. \ref{gr}(a). 
On the other hand, the intramolecular distance between A and B, that is $l$ in Fig. \ref{packing}, is shown as the first peak in Fig. \ref{gr}(b). 
We plotted the dependence of the distance $R$ and the length $l$ on the interaction potential $a_{\rm AA}$  in Fig. \ref{ratio}(a).
Figure \ref{ratio}(a)  indicates that the dependence of the length $l (a_{\rm AA})$ on $a_{\rm AA}$ is weaker than that for $R$.
We also plotted the dependence of  $1/(R^2 l )$ on the interaction potential $a_{\rm AA}$ in Fig. \ref{ratio}(b).

Here we consider the packing parameter introduced by Israelachvili\cite{05Nakamura,76Israelachvili,92Israelachvili}.
According to this description, the hexagonal and micellar phases correspond to $1/2 \le p \le 1/3$ and $p \le 1/3$, respectively. 
The packing parameter $p$ is considered to be proportional to  $1/(R^2 l )$. 
From Fig. \ref{ratio}(b), $p \propto 1/(R^2 l ) $ decreases as $a_{\rm AA}$ increases.
Therefore, it is found that A-B dimers form cylinders for small $a_{\rm AA}$ values, and that AB dimers modify their shape from cylinders to cones by increasing the distance between the neighboring head groups as $a_{\rm AA}$ increases.

Lastly, we compare the present results with our previous work\cite{05Nakamura}, which investigated the dependence of the phase structure on $a_{\rm AW}$.
In that previous work, we showed that the A-B dimer modifies its shape with changing hydrophilicity $a_{\rm AW}$.
It is intuitively expected that the shape of the A-B dimer would depend more strongly on the parameter $a_{\rm AA} $ than the parameter $a_{\rm AW}$, because $a_{\rm AA}$ affects the distance between the hydrophilic groups more directly than $a_{\rm AW}$.
However, contrary to our expectation, the present simulation results show that the dependence of the A-B dimer's shape on $a_{\rm AA}$ is weaker than that on $a_{\rm AW}$.
Based on this result, we conclude that the solvent water $W$ plays an important role in determining the phase structure of the A-B dimers in aggregation.

\begin{table}
\begin{center}
\begin{tabular}{ll}
\begin{minipage}{60mm}
\begin{tabular}{c|ccc}   \hline
$a_{ij}$  & ~~W~~ & ~~A~~ & ~~B~~ \\ \hline
\ \ \ W \ \ \ & 25 &  0    & 50 \\
\ \ \ A \ \ \ &  0  & $a_{\rm AA}$ & 30 \\
\ \ \ B \ \ \ & 50 & 30 & 25  \\  \hline
\end{tabular} 
\caption{Table of coefficients $a_{ij}$ showing dependence on particle type for particles $i$ and $j$, where W is a water particle, A is a hydrophilic particle, and B is a hydrophobic particle. 
By varying the coefficient $a_{\rm AA}$ between adjacent particles A and A, $a_{\rm AA} = 15, 25, 40$ and $250,$ the dependence of the phase structure on the A-A interaction potential can be clarified.  \label{aij} }
\end{minipage}

\hspace*{5mm}
\begin{minipage}{60mm}
\begin{figure}[H]
\begin{center}
\includegraphics[height=2cm]{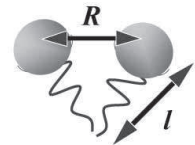}
\end{center}
\caption{Schematic diagram of two adjacent molecules.
A gray ball and a twisting black line are used to denote the hydrophilic and hydrophobic parts, respectively, of an amphiphilic molecule. 
The parameter $R$ is the distance between the hydrophilic groups, 
and $l$ is the ``maximum effective length" of the hydrophobic tail. 
\label{packing} }
\end{figure}
\end{minipage}
\end{tabular}
\end{center}
\end{table}
 
\section{Conclusion}\label{conc}
We demonstrated the dependence of the A-B dimer's aggregated structure in solvent water $W$ by varying the A-A interaction potential $a_{\rm AA}$ in a dissipative particle dynamics simulation.
The present simulation and the previous results\cite{05Nakamura} show that the $a_{\rm AA}$-dependence of the A-B dimer's shape is weaker than the $a_{\rm AW}$-dependence.
Therefore, it is concluded that the solvent water $W$ plays an important role in forming the phase structure of the A-B dimers in aggregation.

\section*{Acknowledgments}  
This research was supported by the Ministry of Education, Culture, Sports, 
Science and Technology of Japan, Grant-in-Aid for Scientific Research (C), 2005, No.17540384.


\begin{thebibliography}{19}
\bibitem{04Nakamura}
Nakamura, H. 2004,  {\it Molecular Simulation} {\bf 30}, 941--945.

\bibitem{05Nakamura}
Nakamura, H. \&  Tamura, Y. 2005, {\it Comp. Phys. Com.\/} {\bf 169}, 139--143.

\bibitem{83Mitchell}
Mitchell, D. J., Tiddy, G. J. T., Waring, L. , Bostock,T.  
\& MacDonald,  M. P.  1983, {\it J. Chem. Soc., Faraday Trans. 1} {\bf 79}, 975--1000.

\bibitem{99Jury}
Jury,S., Bladon, P., Cates, M., Krishna, S., Hagen, M., Ruddock, N. \& Warren P. 
1999, {\it Phys. Chem. Chem. Phys. } {\bf 1}, 2051--2056.

\bibitem{97Groot}
Groot, R. D. \& Warren, P. B. 1997, {\it J. Chem. Phys.} {\bf 107}, 4423--4435.


\bibitem{76Israelachvili}
Israelachvili,J. N., Mitchell, D.~J. \&  Ninham, B.~W. 1976, {\it J. Chem. Soc. Faraday Trans. II } 
{\bf 72}, 1525--1568.

\bibitem{92Israelachvili}
Israelachvili, J. N. 1992  {\em {I}ntermolecular and {S}urface {F}orces} (Academic Press, London) 2nd ed.


\end{thebibliography}
\end{document}